\documentclass[12pt]{amsart}

\title[a semiclassically entangled puzzle]{A semiclassically entangled puzzle}

\author[Pedro de M. Rios]{Pedro de M. Rios}

\address{Department of Mathematics, University of California, Berkeley, CA, 94720-3840, USA.}
\email{prios@math.berkeley.edu}

 1

\def\d{{\partial}} 

\begin{document}

\thispagestyle{empty}

\begin{abstract}

For a maximally entangled eigenstate of a system of two non-interacting 
identical one dimensional harmonic oscilators, at the semiclassical level, 
it is not obviously true that a nonlinear interaction with one of 
the subsystems leaves the reduced semiclassical Wigner 
function of the other subsystem unaffected. Once stated, we advance some 
wild speculations regarding this seeming puzzle.   

\end{abstract}

\maketitle 

\thispagestyle{empty}

\section{introduction}

When Einstein, Podolsky and Rosen \cite{EPR}  explored the ``strange'' behaviour of entangled  
states, their main goal was  to set forth an argument  
concerning the ontology of quantum mechanics. Only quite later, did Bell \cite{B} show that some of the 
discussion had a definite empirical bearing, namely, on matters related to the nonlocal 
character of quantum mechanics. After the empirical confirmation of this latter \cite{ADR},   
emphasis has been placed on the ``peaceful coexistence'' \cite{Shim} of this kind of nonlocality and 
the partial order of spacetime events derived from the principle of relativity \cite{E}.  

In terms of the  Wigner \cite{Wig} function $W$, the cartesian Weyl \cite{W} 
representation of the density operator $\hat{\rho}$ 
as a real function on euclidean phase space,    
this question is addressed in the 
following way: let $x_1$ and $x_2$ denote the phase 
space points of two subsystems whose total system is in an entangled eigenstate of the hamiltonian $\hat{H}_1 + \hat{H}_2$ , 
with corresponding Wigner function $W_0(x_1,x_2)$. 
Also, let $\hat{H}_2'$ be a new hamiltonian, whose Weyl symbol $H_2'$ generates a flow 
$x_2(0) \to x_2(t)$, and let $W_t$ denote the evolution of $W_0$ effected by this new hamiltonian.  
Further: let $W^1_0(x_1) \equiv \int W_0(x_1,x_2)dx_2$ and $W^1_t(x_1) \equiv \int W_t(x_1,x_2)dx_2$ be the 
corresponding reduced Wigner functions, which encode 
all the empirical information of subsystem $1$ alone, and denote their difference by $\delta W^1(x_1,t)$.  
 
In this context, conservation of the relativistic partial order of events is implied by  
\begin{equation} \label{lin} 
\delta W^1 \equiv  \ 0 \ , 
\end{equation}
which is always true in full quantum mechanics $(\hbar = 1)$, due to its linear structure \cite{G}. 
In a classical setting  $(\hbar = 0)$, 
on the other hand, a sufficient condition for the validity of (\ref{lin}) is given by  
\begin{equation} \label{folk} 
W_0(x_1,x_2(0)) \longrightarrow W_t(x_1,x_2(t)) = W_0(x_1,x_2(0)) \ .
\end{equation} 

Years ago, Heller \cite{Hel} noticed that the generic expression corresponding to (\ref{folk}) 
is not always valid at the semiclassical level $(\hbar \to 0^+)$, a fact which 
has recently been reinterpreted in a more geometrical way \cite{RO}. 
Accordingly, for a generic semiclassical Wigner function 
$\mathcal{W}$, (\ref{folk}) is generally true only if $H_2'$ is quadractic.  
For a nonlinear interaction, (\ref{folk}) is valid only for points $x_2$ close to the Lagrangian leaf 
corresponding to the classical limit of $W$, whenever its semiclassical expression $\mathcal{W}$ is an 
oscillatory function of $x_2$, as is usually the case for pure states \cite{BB}\cite{Hel}\cite{RO}.  
 
The purpose of this letter is to state the following puzzle: at the semiclassical level, 
it is {\it not} obvious whether (\ref{lin}) is valid in general. 
To see this puzzle, we focus on a particularly 
simple semiclassically entangled state, namely the (anti)symmetric state of two identical $1$-dimensional harmonic oscilators, under a small nonlinear interaction $H_2'$. 
We apply the results of \cite{RO} to ask whether, in a suitable limit, \linebreak(\ref{lin}) might not hold. 
Finally, we emphatically stress the highly speculative nature of this letter. 

\section{the setup}   

We study the semiclassical dynamics of an eigenstate  
of $\hat{H} = \hat{H}_1 + \hat{H}_2$ , where $\hat{H}_i = (\hat{p}_i^2 + q_i^2)/2$, \linebreak and of $\hat{P}_{12} : 1 \leftrightarrow 2$. 
This state is:   
$\sqrt{2}|\Psi\rangle = |e_m,1\rangle\otimes|e_n,2\rangle \pm |e_n,1\rangle\otimes|e_m,2\rangle$ , where the $\pm 1$ refers to 
the eigenvalue of $\hat{P}_{12}$ and where $e_{m+n}=e_m+e_n$ is the 
eigenvalue of $\hat{H}$ , with $e_m = (m+1/2)\hbar$ , $e_n = (n+1/2)\hbar$ and where the energy difference is $\lambda = e_m-e_n =l\hbar$, with $m-n=l>0$.

The corresponding  Wigner function factors as 
$2W_{m+n}(x_1,x_2) = W_m(x_1)W_n(x_2) + W_n(x_1)W_m(x_2) 
\pm M_m^n(x_1)M_n^m(x_2) \pm  M_n^m(x_1)M_m^n(x_2)$ , where $M_m^n$ is the Moyal \cite{M} function, 
or cartesian Weyl representation of the transition operator $|e_m\rangle\langle e_n|$, satisfying $M_m^n = \overline{M_n^m}$. 
We write this function in polar $\{x=(r,\theta)\}$ 
coordinares  but emphasize that we are not considering the  action-angle Weyl representation, 
which is not equivalent  \cite{Ber}. 
In taking the semiclassical limit \footnote{There are different phenemenological ways of approaching 
the classical limit. The one considered here, large quantum numbers or small effective Planck's constant, 
should not be identified with the limit of many degrees of freedom, 
or large complexity, for which most present theories of decoherence apply.}, 
we formally let $\hbar \to 0^+$ . This is equivalent to letting $
m , n \to \infty$ , while keeping $e_m, e_n < \infty$ , or to varying an ``effective'' $\hbar$.     

Our analysis is based on the semiclassical expression for the Moyal function which is not corrected by 
uniform approximations on the caustics. Ours is a 
particularly simple case of the general expression carefully treated in \cite{OdA2} and is given by 
(we use curly letters to indicate semiclassical expressions): 
\begin{equation} \label{Moy} 
\mathcal{M}_m^n(x) \ = \ \frac{e^{il \theta }cos(\phi_{m,n}(r)/\hbar -\pi/4)}
{\sqrt{\pi^3\hbar /2} \ D_{m,n}(r)} \ , 
\end{equation} 
where  $\phi_{m,n}(r)$ is half of the symplectic area between the circle with radius $\sqrt{(2m + 1)\hbar}$ 
centered at the origin and the circle with radius 
$\sqrt{(2n + 1)\hbar}$ centered at $(2r , \theta)$ , and $D_{m,n}(r) = |\dot{x}_m^+\wedge\dot{x}_n^-|^{1/2}$,   
where $\dot{x}_m^+$ , $\dot{x}_n^-$ are the phase space velocity vectors of the intersecting circles. 
These intersections are real as long as $r_{-} < r < r_{+}$, 
where $r_{\pm} = (\sqrt{(2m + 1)\hbar} \pm  \sqrt{(2n + 1)\hbar})/2$ ,  and 
the denominator goes to zero when the circles become tangent, i.e. $r=r_{\pm}$ , 
which are the caustic lines. Outside this ring, $\mathcal{M}_m^n \equiv 0$ , 
in this crude expression. Naturally, the corresponding formula for $\mathcal{W}_m$ is obtained from (\ref{Moy})  
with $m=n$. By performing uniform approximations as the intersection points coalesce, 
$\mathcal{M}_m^n$  can be expressed 
in terms of Airy functions which do not blow up and decay exponentially outside the ring. 
Still using (\ref{Moy}), the  sum of the cross terms in $\mathcal{W}_{m+n}$ is expressed as   
\begin{equation} \label{ct} 
Re\{ \mathcal{M}_m^n(x_1)\mathcal{M}_n^m(x_2) \} \ = 
\ \frac{cos(l (\theta_1 - \theta_2))cos(\phi_{m,n}(r_1)/\hbar -\pi/4)cos(\phi_{m,n}(r_2)/\hbar -\pi/4)}{(\pi^3\hbar /2) 
D_{m,n}(r_1)D_{m,n}(r_2)} \ .
\end{equation}
Of course, $\int Re\{ \mathcal{M}_m^n(x_1)\mathcal{M}_n^m(x_2)\} dx_2 = 0$ and so 
$\mathcal{W}^1_{m+n} \equiv (\mathcal{W}_{m} + \mathcal{W}_{n})/2$ , as it should be.  

\section{the puzzle} 

The seeming puzzle to be laid forth below follows from the fact that 
$\mathcal{M}_m^n$ is an {\it oscillatory} function, that is, 
(\ref{Moy}) has an essential singularity at $\hbar = 0$. This is the reason why  
(\ref{folk}) fails to hold when $\hbar \to 0^+$  
as can be seen by directly plugging (\ref{Moy}) into the Moyal bracket.  
Its successive derivatives then bring negative powers of $\hbar$ which alter the usual 
$\hbar$-expansion obtained for smooth functions. 
Recently \cite{RO}, a simple geometrical prescription for a more correct 
semiclassical propagation of such functions was obtained, 
in agreement with another old result derived by Berry and Balazs \cite{BB} 
according to which one should propagate the whole family of curves, 
or lagrangian submanifolds, associated with a quantum state and then 
re-evaluate the semiclassical expressions accordingly. 

The new result obtained in  \cite{RO} somehow simplifies this procedure  
by showing that, equivalently, one can obtain a more correct 
semiclassical propagation of an oscillatory function like $\mathcal{M}_m^n(x)$ 
by classically propagating the {\it tips} of the 
chord centered on $x$ instead of classically propagating the argument $x$ itself, 
as in the Liouville case (\ref{folk}). 
Of course, Lioville propagation is still semiclassically correct for smooth functions, 
when all chords are null, or if the flow is 
linear, as discussed in more detail in \cite{RO}. Since $\mathcal{M}_m^n(x)$ is not smooth, 
we might expect surprises when under nonlinear interactions. 
And these might be related, in some way or another, to the old strange fact that for 
generic pure-state Wigner functions under nonquadractic hamiltonians,   
if the interaction is first performed in the full quantum regime and then the classical limit is 
taken, the final result generally appears to be rather different from first taking the classical limit and then interacting. 

Therefore, let us now suppose that subsystem $2$ interacts nonlinearly for an interval of time $t$. 
For simplicity, we state the puzzle when the interaction is 
either cubic $H_2' = \alpha q_2^3/3$ or quartic $H_2' = \alpha q_2^4/4$  
and denote by $\epsilon = \alpha t$ the strength of this interaction. 
After rewriting (\ref{ct}) as a sum of oscillatory functions of 
$\phi_{m,n}^{\pm}(r_2, \theta_2) = \phi_{m,n}(r_2) \pm \lambda\theta_2$ ,  
it is not too difficult to see that, within the stationary phase regime $(\hbar \to 0^+)$, 
the effect of a small interaction is, in a first approximation, the equivalent to having 
only phase shifts $\phi_{m,n}^{\pm}(r_2, \theta_2) \to \phi_{m,n}^{\pm}(r_2, \theta_2) + \delta\phi_{m,n}^{\pm}(r_2,\theta_2)$. 
 
More precisely, to get these phase shifts  
we first apply equation (16) in \cite{RO} to the chord $\xi_{m,n}^{\pm} = -J(\d \phi_{m,n}^{\pm}(x_2)/\d x_2)$, 
centered on $x_2$, and get the ``midpoint phase difference'' $\tilde{\phi}_{m,n}^{\pm}(\tilde{x}_2^t) - \phi_{m,n}^{\pm}(x_2)$, 
where $\tilde{x}_2^t$ is the midpoint 
of the hamiltonian flow of the tips of the chord $\xi$. Then we note that, for small $\epsilon$, the total phase difference acquired 
for the hamiltonian flow of $x_2$ is approximately half of $\tilde{\phi}_{m,n}^{\pm}(\tilde{x}_2^t) - \phi_{m,n}^{\pm}(x_2)$. 
The amplitude being approximately covariant under the hamiltonian flow, for small $\epsilon$, this determines the looked-for phase shifts as   
$$
\delta\phi_{m,n}^{\pm}(r_2,\theta_2) \approx 
(2\epsilon/3)\{sin(\theta_2)[e_{m+n} - (\lambda/2r_2)^2 - r_2^2]^{1/2} \pm cos(\theta_2)[\lambda/2r_2]\}^3 \ ,  
$$
for the cubic case, while  for the quartic case we get  
$$
\delta\phi_{m,n}^{\pm}(r_2,\theta_2) \approx  (2\epsilon) r_2cos(\theta_2)\{sin(\theta_2)[e_{m+n} - (\lambda/2r_2)^2 - r_2^2]^{1/2} \pm cos(\theta_2)[\lambda/2r_2]\}^3  \ . 
$$

The next step is to compute $\int Re\{ \mathcal{M}_m^n(x_1)\mathcal{M}_n^m(x_2)\} dx_2$ 
and check if it vanishes, as before. 
But note that the phase shifts have broken the symmetry of $Re\{ \mathcal{M}_m^n(x_1)\mathcal{M}_n^m(x_2)\}$ 
with respect to $\theta_2$ and it is not obvious now whether the integral vanishes or not. In fact, we 
must now perform the new integration $\int Re\{ \mathcal{M}_m^n(x_1)\mathcal{M}_n^m(x_2)\} dx_2$ by stationary 
phase and it is not difficult to 
see that the critical points of $\phi_{m,n}^{\pm}(r_2, \theta_2) + \delta\phi_{m,n}^{\pm}(r_2,\theta_2)$ 
are generally complex. 
This means that we must generally perform a 2-dimensional steepest descent evaluation of the integral, 
carefully   deforming the whole ring into the complex 2-plane, counting all the significant contributions, 
adding them all up...  
This is a rather delicate computation, but it is not unreasonable to suppose that, instead of zero, 
we should at least get an exponentially small result \cite{K}, as in a ``tunneling'' effect \footnote{
For a much simpler analogy, consider the integral 
$\mathcal{I}$ $=$ $ \int_0^{2\pi} cos(l\theta) d\theta$ . Writing 
$l\theta=\phi(\theta)/\hbar$ , where $\phi(\theta)=\lambda\theta$, 
then under the phase shift $\phi \to \tilde{\phi}=\phi+\delta\phi$, 
where $\delta\phi=\epsilon \ \theta^3/3$ , we have that 
$\mathcal{I} \to \widetilde{\mathcal{I}}$ is asymptotically $(\hbar \to 0^{+})$ 
mapped into the Airy integral, that is, $\widetilde{\mathcal{I}}$ $\approx$ 
$\pi(\hbar/\epsilon)^{1/3} \mathcal{A}_i(y)$ , where $y=\lambda\epsilon^{-1/3}\hbar^{-2/3}$.}. 

This being an asymptotic analysis, however, we still need to make sure whether the result of this 
stationary phase computation implies that $\delta\mathcal{W}^1$ is {\it really} different from zero, 
to lowest order of $\hbar$ (remember that (\ref{Moy}) is only the first term of an $\hbar$-expansion), 
in some region of the positive $(\lambda,\epsilon,\hbar)$-space close to the $\hbar = 0$ plane.  
Also in this region, we must analyse the behaviour of $\int \mathcal{W}_{m}(x_2) dx_2$ and 
$\int \mathcal{W}_{n}(x_2) dx_2$ under  
similar phase shifts  to find out how unitary such a propagation of 
$\mathcal{W}^1$ actually is 
(note that $\int\int Re\{ \mathcal{M}_m^n(x_1)\mathcal{M}_n^m(x_2)\} dx_1dx_2$ vanishes). 
Then we must compare all the above results and further analyse and interpret them 
(for instance, in possible relation to Stokes phenomenon).  

Finally, to properly account for stronger interactions ($\epsilon$ not so small), 
closer resonance ($\lambda$ small), 
or to actually get more  precise estimates for $\delta\mathcal{W}^1$ 
(any $\epsilon$ or $\lambda$), we must leave the crude expression (\ref{Moy})  
and proceed much more carefully with the uniform approximations (this is the really difficult part), 
since non-Liouville propagation of the amplitude is so much more relevant near caustics.  

It is very important to emphasize that
any possible nontrivial answer for $\delta\mathcal{W}^1$ 
would only be an \linebreak asymptotic one. Such a possibility  seems   
due to the $\theta_2$-symmetry breaking of the whole semiclassical Wigner function $\mathcal{W}_{m+n}(x_1,x_2)$, 
effected by the nonlinear interaction with $H_2'$ , and the consequences of     
this symmetry breaking should become more explicit as  
the amplitude corrections are included. 
But, actually, this is precisely the very hard crucial point that needs to be fully investigated. 
In other words, it is quite likely that 
only a careful (and very difficult) analysis based on more accurate 
semiclassical approximations can produce clear answers, 
even because the accuracy of the chord expression (\ref{Moy}), from 
which we started, is in fact worse than exponential.  

Then, {\it if} such a possibility is indeed correct, we could 
conjecture there would be an optimal way of approaching the classical limit, 
in the positive $(\lambda,\epsilon,\hbar)$-space, leading to a maximal $\delta\mathcal{W}^1$. 
Actually, combined with the singular nature of the semiclassical limit, 
the presence of so many independent parameters is what seems to open this problem in such a manner. 
And for more degrees of freedom, issues from the classical (non)integrability of a subsystem, 
with respect to the nonlinear interaction, might well play a further significant role.    
However, it is clear that nothing in the previous discussion proves or even suggests that a nontrivial 
$\delta\mathcal{W}^1$ can probably happen. In this letter, 
we have merely pointed out to an apparent possibility that this somehow might come to pass.  

\section{speculations}  

Final words on this puzzle from the phenomenological, empirical and philosophical points of view.  

Phenomenologically, how could it ever be possible for (\ref{lin}) to fail at all, 
since it is valid in both the classical and the full quantum regimes 
\footnote{Examples of phenomena that appear in one or the other but not in both regimes are easier to come by. 
On the other hand, the non-geometrical phenomenon of the glory \cite{Nu} can be seen as a phenomenon that vanishes 
both ``classically'' (ray regime) and ``fully quantum'' (long wave regime) 
but not ``semiclassically'' (short wave asymptotics).} ?   
Well, we could argue that (\ref{lin}) is valid in both regimes for different reasons. 
In the classical case, (\ref{folk}) translates the fact that no quantum correlations have ``survived''. 
But this can be seen as an effect of the correlations becoming ``random'', 
which is best expressed by the oscillations in the Wigner function becoming so high that they 
average to zero everywhere but on the classical leaf.  
In the full quantum case, by contrast, the correlations 
are fully structured, in other words, they have a definite ``form'', which is usually expressed by the concept of 
a definite normalized vector in the Hilbert space. Such a vector can only be preserved by linear transformations, 
from which (\ref{lin}) follows. 
Therefore, the puzzle ammounts to a question of whether, at the semiclassical level, 
the linear structure of quantum mechanics could feel some disturbing effects of a nonlinear classical dynamics 
and, if so, how they could show up 
\footnote{In a certain way, the two important factors for a Bell's telephone, namely the 
distinction between pure and mixed states and between linear and nonlinear evolution, show up  
in the semiclassical propagation of Wigner functions.}. 

Empirically, if a suitable limit where $\delta\mathcal{W}^1 \neq 0$ could ever be found, 
would this mean the possibility of faster-than-light-communication signals? 
Here, we should stress that these sub-hypothetical ``{\it falico}'' signals 
would be fundamentally threshold phenomena and  
so a deeper understanding of this threshold between the quantum and the classical regimes could be required. 
In other words, the very existence of stable semiclassically entangled states would be at question. 
On the other hand, even if granted their existence for some special systems, 
in a suitable limit of nontrivial $\delta\mathcal{W}^1$, 
it could still be a formidable, pehaps nearly impossible  
task to actually measure such signals, even though some direct measurements of Wigner functions 
(if ever needed) have already been claimed \cite{LD}. 

Philosophically, what would be at stake here is, among other things, the principle of relativity. 
But in this respect, it seems interesting to point out the usually overlooked fact that, 
as originally formulated \cite{E}, 
the relativistic partial order presupposes events which are correlated by an exchange of energy. 
In other words, which are correlated by ``material'' causes. Such is not the case for 
the quantum correlations treated here, which could pehaps be described as ``formal'' correlations.   
Maybe, these two kinds of correlations being essentially distinct, their principles need not be the same. 

\newpage 

\noindent
{\it Acknowledgements}: Speculations supported by CNPq via a postdoctoral fellowship to UC Berkeley. 
I thank Alan Weinstein, Jair Koiller, Robert Littlejohn, A. Ozorio de Almeida and Michael Berry 
for some pertinent, though not necessarily positive, comments 
(the responsability for these speculations is entirely mine). 
This letter is not under consideration for publication in any other way, or elsewhere.

\end{document}